\documentclass[sigconf]{acmart}

\usepackage{booktabs} % For formal tables

\setcopyright{rightsretained}
\def \xx {{\mathbf{x}}}
\def \ww {{\mathbf{w}}}
\def \WW {{\mathbf{W}}}
\def \ee {{\mathbf{e}}}
\def \BB {{\mathbf{B}}}
\def \YY {{\mathbf{Y}}}
\def \vv {{\mathbf{v}}} 
\def \uu {{\mathbf{u}}}
\def \pp {{\mathbf{p}}}
\def \XX {{\mathbf{X}}}
\def \UU {{\mathbf{U}}}
\def \VV {{\mathbf{V}}}
\def \PP {{\mathbf{P}}}
\def \BB {{\mathbf{B}}}

\acmDOI{10.475/123_4}

\acmISBN{123-4567-24-567/08/06}

\acmConference[ACM'18]{ACM Mobile AI conference}{Dec 2017}{
USA} 
\acmYear{2018}
\copyrightyear{2016}

\acmArticle{4}
\acmPrice{15.00}

\begin{document}
\title{Science Driven Innovations Powering Mobile Product: Cloud AI vs. Device AI Solutions on Smart Device}
%\titlenote{Produces the permission block, and copyright information}
%\subtitle{Extended Abstract}
%\subtitlenote{The full version of the author's guide is available as
 % \texttt{acmart.pdf} document}
%\author{Deguang Kong, Jimmy Yang, Konstantin Shmakov and Chiranjeevi Devi}

\author{Deguang Kong}
%\author{Deguang Kong, Jimmy Yang, Konstantin Shmakov and}
%\authornote{Dr.~Trovato insisted his name be first.}
\affiliation{%
  \institution{Yahoo Research}
  \streetaddress{701 1st Ave, Sunnyvale, California,  94089}
}
\email{ doogkong@gmail.com}


\begin{abstract}
Recent years have witnessed the increasing popularity of mobile devices (such as iphone) due to the convenience that it brings to human lives.  On one hand, rich user profiling and behavior data (including per-app level, app-interaction level and system-interaction level) from heterogeneous information sources make it possible to provide much better services (such as recommendation, advertisement targeting) to customers, which further drives revenue from understanding users' behaviors and improving user' engagement. In order to delight the customers, intelligent personal assistants (such as Amazon Alexa, Google Home and Google Now) are highly desirable to provide real-time audio, video and image recognition, natural language understanding, comfortable user interaction interface, satisfactory recommendation and effective advertisement targeting. 

This paper presents the research efforts we have conducted on mobile devices which aim to provide much smarter and more convenient services by leveraging statistics and big data science, machine learning and deep learning, user modeling and marketing techniques to bring in significant user growth and user engagement and satisfactions (and happiness) on mobile devices.  The developed new features are built at either cloud side or device side, harmonically working together to enhance the current service with the purpose of increasing users' happiness. We illustrate how we design these new features from system and algorithm perspective using different case studies, through which one can easily understand how science driven innovations help to provide much better service in technology and bring more revenue liftup in business.  In the meantime, these research efforts have clear scientific contributions and published in top venues, which are playing more and more important roles for mobile AI products.
\end{abstract}


\keywords{engagement, growth,  big data science,  recommender, targeting, forecasting, effectiveness,  deep,  convolution, GoogLeNet, optimization, embedding, LSTM, natural language, malicious, adversarial, privacy
}

\maketitle

\section{introduction}
Artificial intelligence (AI) is the technology driving the new revolution. AI on cloud is running AI algorithms on cloud side, and it has demonstrated overwhelming performance in image recognition, speech recognition, and video understanding, etc.  AI on device is running powerful AI algorithms on the device.  Due to the limited resources on device side, AI algorithm must be more economic and efficient to satisfy real-time serving requirement.  

New challenges come with the exponentially growing markets of mobile device applications. We need to address many new problems, for example, diversified app markets, heterogeneous  users' behaviors and limited computational resources,  in order to provide better service and improve user engagement on mobile devices. AI is playing more and more important roles in mobile product, which is highly demanded by customers to provide more intelligent, smart and convenient services.  In the paper next, we will show our research efforts towards building smarter and more convenient mobile systems in the following aspects:

\begin{itemize}

\item Cloud-AI solutions:  The majority of computing and learning tasks are performed at cloud side, which provides rich resources and computational powers for computing. Cloud is necessary and very helpful for pooling of big data and training huge amount of history observations using sophisticated machine/deep learning and parameter tuning strategies. 
	
	\begin{itemize}
	
	\item Mobile App Understanding: Risk Assessment and Malicious App detection 
	
	\item Marketing for mobile app User Growth: Attraction new users, retention, campaign effectiveness analysis
	
	\item Mobile App User engagement:  User modeling, profiling and Recommendation
	
	\item Mobile App Monetization:  native advertisement serving, advertisement targeting, bid optimization and pricing
	
         \end{itemize}
	
\item Device AI solutions: The major learning and recognition tasks are performed at device side.   Running AI on device has brought many advantages, such as immediate response, enhanced reliability, increased privacy, and efficient use~\cite{li2015emod} of network bandwidth. AI inference algorithm can be mainly running on mobile devices. 

	\begin{itemize}
	
	\item Image Recognition and Image Privacy on Mobile Devices
		
        \item Deep learning model compression on mobile devices

	\end{itemize}
	
\item Cloud AI and Device AI interactions:  Some parts of learning and recognition tasks are performed at device side while others are put on cloud side.  One particular example is personal assistant, such as Amazon Alexa and Google Home.  In particular, AI inference running entirely in the cloud will have issues for real-time serving that are usually latency-sensitive and mission-critical (e.g., autonomous driving).  The real-world system is expected to benefit from the cloud-side sophisticated training and high-performance device processing and inference~\cite{li2016deepcham}, leading to the best overall system performance.

	\begin{itemize}
	\item Personal Assistant Engine on Mobile Device
	\begin{itemize}
	\item Dialog System
	\item Speech recognition 
	\item Speech synthesis
	\item NLP understanding
	\item Chatbot
	\item Recommender system
	\end{itemize}
	\end{itemize}
	
\end{itemize}

We provide scientific leaderships and provide practical solutions to solving these challenging problems because we strongly believe that, pure engineering is not enough in practice. These techniques will be very helpful for addressing the challenging problems in solving mobile big data science and AI problems in real world if (a big ``if'') the technology can be accurate, robust and scalable enough. We need research breakthrough to improve the state-of-the-art and we have diligently worked on them. Fortunately, some of them have been graduated into product.

\begin{figure}[t]
	\centering
	\includegraphics[height=1.8in,width=0.35\textwidth]{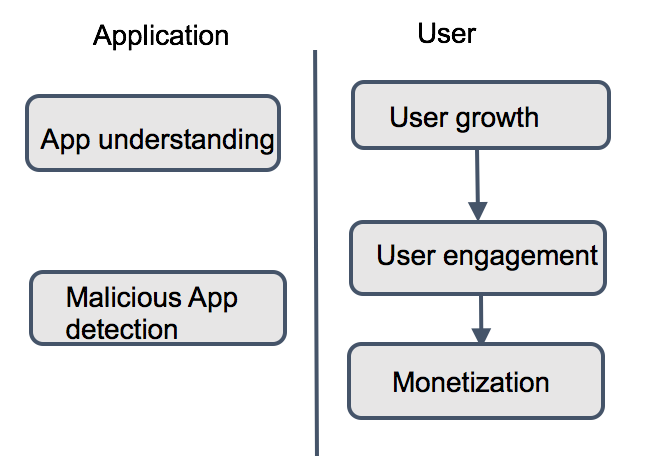}
		\caption{\small{Overview of cloud AI solution: (a) left panel: from application perspective, understanding apps and detection of malicious apps; (b) right panel: user growth, user engagement and monetization. 	}
	}
	\label{fig:cloud-ai}
\end{figure}

\section{Cloud AI solution} 

In this section, we provide cloud-AI solutions from the following four perspectives: 
\begin{itemize}
	\item Mobile App Understanding: Risk Assessment and Malicious App detection 
	
	\item Marketing for mobile app User Growth: Attraction new users, retention, Campaign effectiveness analysis
	
	\item Mobile App User engagement:  User modeling, profiling and Recommendation
	
	\item Mobile App Monetization:  advertisement serving, advertisement targeting, bid optimization and pricing
	
\end{itemize}
In particular, Fig.\ref{fig:cloud-ai} shows an overview of the framework.

\subsection{Mobile App Understanding}
%%%====
\begin{figure}[t]
	\centering
	\includegraphics[height=1.5in,width=0.45\textwidth]{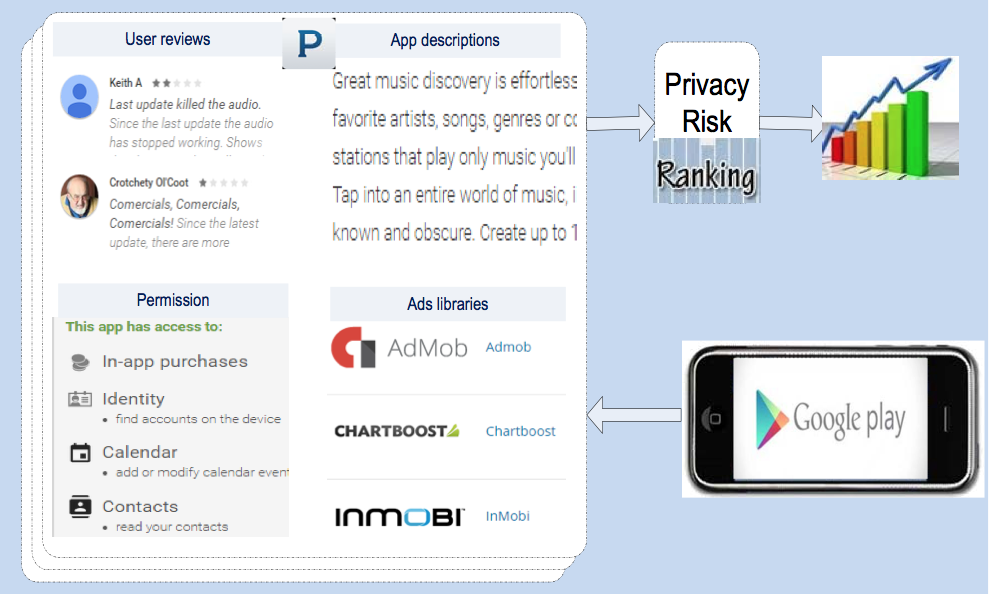}
		\caption{\small{Ranking the risk of mobile apps using multi-modal features such as descriptions, user review, 
permission access and ad library. }
	}
	\label{fig:risk}
\end{figure}
%%%=====
\begin{figure*}[t]
	\centering
	\includegraphics[height=2.2in,width=0.9\textwidth]{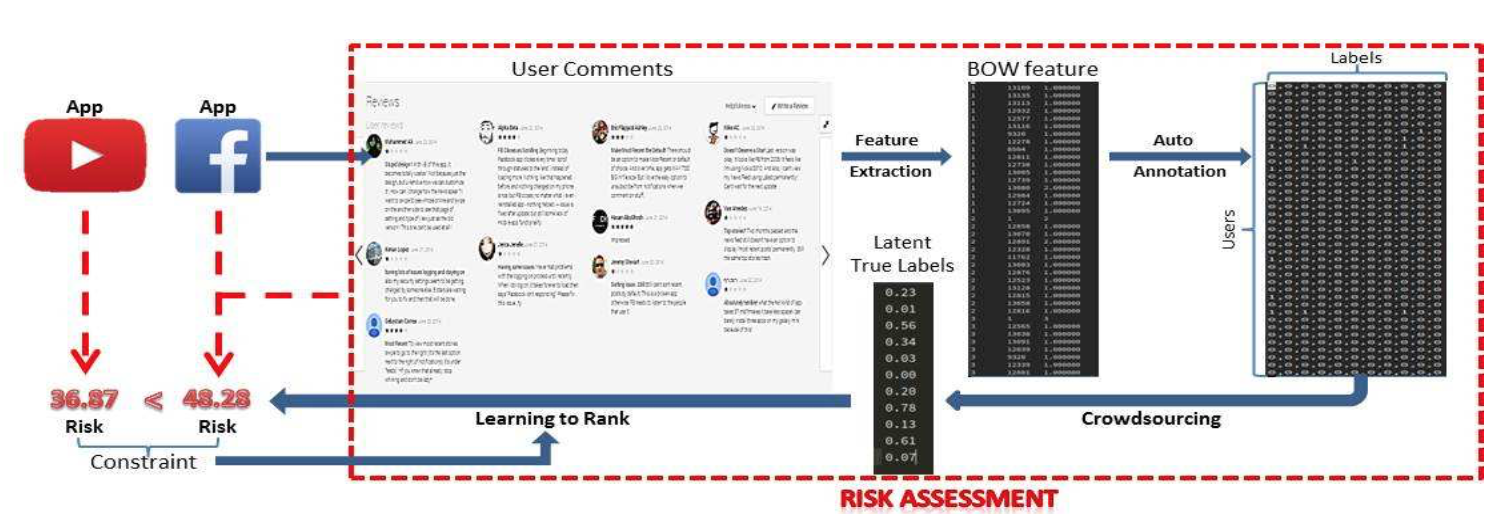}
		\caption{\small{
		Flowchart of risk assessment from user comments. Given an app, the user comments are used to evaluate the risk of apps in two steps: (a) ``crowdsourcing'' is used to accumulate user comments into app-level features to accumulate user comments into app-level features (shown as ``feature extraction'', ``auto annotation'' and ``crowdsourcing''); (b) ``learning to rank'' model is used to predict risk scores by utilizing these latent features, where pairwise constraints are enforced between pairwise apps (shown as relative scores of two apps).
}
	}
	\label{fig:risk-comment}
\end{figure*}
%%%=======
\begin{figure}[t]
	\centering
	\includegraphics[height=1.8in,width=0.45\textwidth]{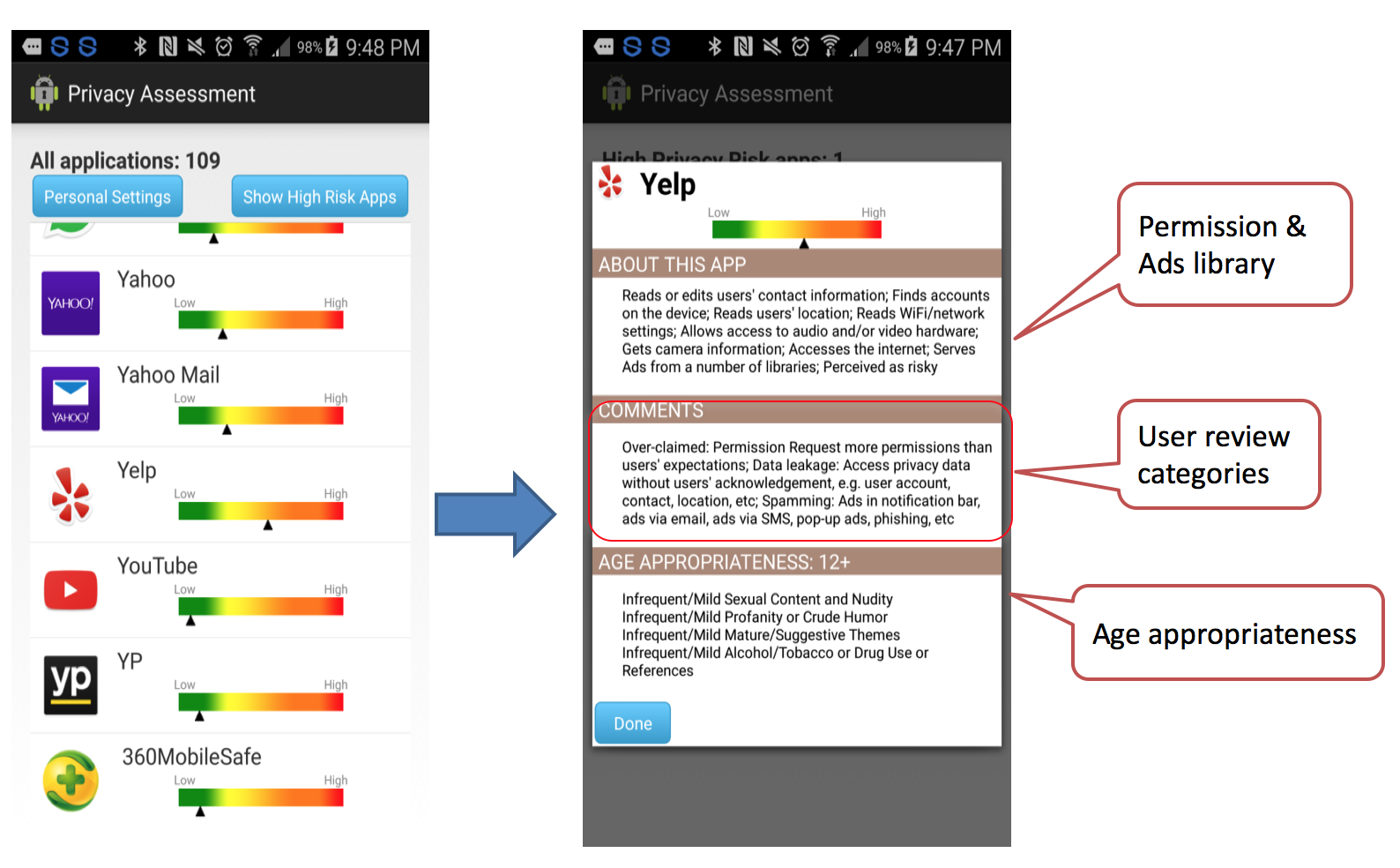}
		\caption{\small{Demonstration of app risk assessment app. }
	}
	\label{fig:app_all}
\end{figure}
%%%=====
A mobile app is a software application developed specifically for the use on smart devices. The mobile can provide the convenience to users to achieve the users' purpose and satisfy users' interest, almost solving everything you may image. For example, one can get news from the news app, one can purchase a Swiss watch while sitting at any office, one can even talk with the lover who is actually far away for thousands of miles. In this section, we will provide the mobile app understanding from several perspectives: 

$\bullet$ App risk assessment

$\bullet$ Malicous app detection 

$\bullet$ App maturity rating

\subsubsection{App Risk Assessment}

$\\$ 
Comparing with traditional software markets, markets like Google Play and Apple Store have lower entry threshold for developers and faster financial payback, hence greatly encouraging more and more developers to invest in this thriving business. Therefore controlling the quality of apps, especially the security risk of them across the whole markets, becomes an important issue to all that involved.  On the other hand, public concerns about privacy issues with online activity and mobile phones are also elevating, demanding a mobile environment with more respect to users' privacy.  In mobile apps, permissions indicate the resources that the apps can access, and thus can be viewed as a privacy indicator.  From users' perspective, the meta data such as user reviews and developer descriptions reflect users' perceptions and developer expectations for the apps, and thus are also correlated with risks of apps.

Our idea is to explore heterogeneous privacy indicators ~\cite{DBLP:conf/sdm/KongJ15}~\cite{ kong-heter} for app risk ranking, which, we believe, is very important to internet company to improve user engagement in mobile platforms (as shown in Fig. \ref{fig:risk}). The risk ranking problem is formulated as a multi-view feature learning problem by exploring group LASSO and exclusive group LASSO techniques~\cite{DBLP:conf/nips/KongF0ND14}, ~\cite{ DBLP:conf/icdm/KongD13},~\cite{DBLP:conf/aaai/KongLLB16} which can automatically select the most discriminant features by considering both inter-view feature competitions, and also intra-view feature correlations.  In particular, we solve the following optimization problem, given feature $\xx_i \in \Re^p$   for each app $i$,  $Y_{ki}$ for label of app $i$ with category label $k$ ($1 \leq k \leq K$), we aim to find the feature weight $\ww^v_k$ for class $k$ regarding $v$-th ($1 \leq v \leq V$) view feature, {\it i.e.,}
\begin{eqnarray}
\min_{\WW \in \Re^{p \times K}} & \sum_{i=1}^n \sum_{k=1}^K \Big(Y_{ki} log \sum_{k=1}^K  \exp^{ \ww^{\top}_k \xx_i} - Y_{ki} \ww^{\top}_k \xx_i \Big) \nonumber \\
 &  + \alpha \sum_{k=1}^K \sum_{v=1}^V \| \ww^v_k\|_2 + \beta  \sum_{k=1}^K \sum_{v=1}^V \| \ww^v_k\|^2_1 
\end{eqnarray}

Correspondingly, we derive an efficient iteratively re-weighted algorithm to tackle the resultant optimization problem, which can handle any group structure, regardless of coherent or exclusive group structures. It demonstrates very good performance in real-world datasets (totally 13, 174 apps, 34, 514 descriptions, 9,986, 568 user reviews and 100 ads libraries).

We also derive a crowdsourcing ranking   approach~\cite{DBLP:conf/ccs/KongCJ15}~\cite{ DBLP:conf/sdm/CenKJS15} (see Fig.\ref{fig:risk-comment}) to rank risk of apps from user comments by combining feature learning and ranking SVM methods, which also provides good solutions in practice. The problem we solve is formalized as:
\begin{eqnarray}
\min_{ \ww \in \Re^p, \theta,  \YY_{\ell} }  & -log \Pr(D_n | \theta, \YY_{\ell}) -log \Pr(\theta)  \nonumber \\
& \lambda \|\ww\|^2 + C \| (\ee - \BB \YY_{\ell} \ww \|^2, 
\end{eqnarray}
based on the maximum a posteriori probability (MAP) estimation, where $\ww$ is the feature weight, $\YY_{\ell}$ is the labels learned
from the feature pre-processing step and $\theta$ is the prior distribution of parameters in crowdsourcing process by aggregating different user reviews, and $log Pr(D_n | \theta, \YY_{\ell})$  gives the likelihood of objective  function given the current parameters while 
$\| (\ee - \BB \YY_{\ell} \ww \|^2$ is the hinge-loss function in SVM ranking.

{\bf Lessons Learned} We do need multi heterogeneous models to find the most discriminant features. User reviews and ads libraries play import roles in understanding the risk of apps except the permissions. A demo system is shown in Fig.\ref{fig:app_all}.

\subsubsection{Malicious App Detection}

\begin{figure}[t]
	\centering
	\includegraphics[height=1.1in,width=0.5\textwidth]{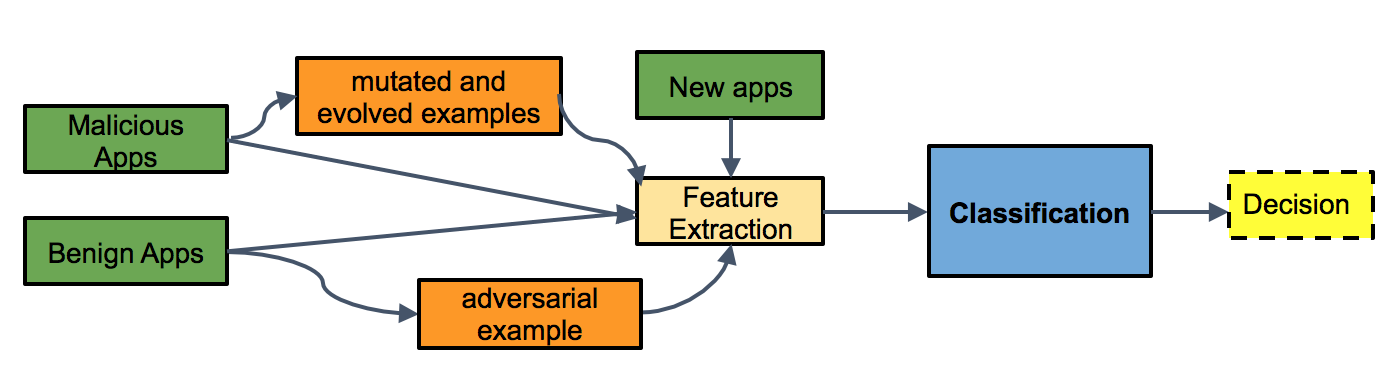}
		\caption{\small{Enhanced malicious app detection process: (i) for benign apps, we generate adversarial examples; (ii) for malicious apps, we generate the mutated and evolved apps.}
	}
	\label{fig:malware}
\end{figure}

\begin{figure}[t]
	\centering
	\includegraphics[height=2in,width=0.4\textwidth]{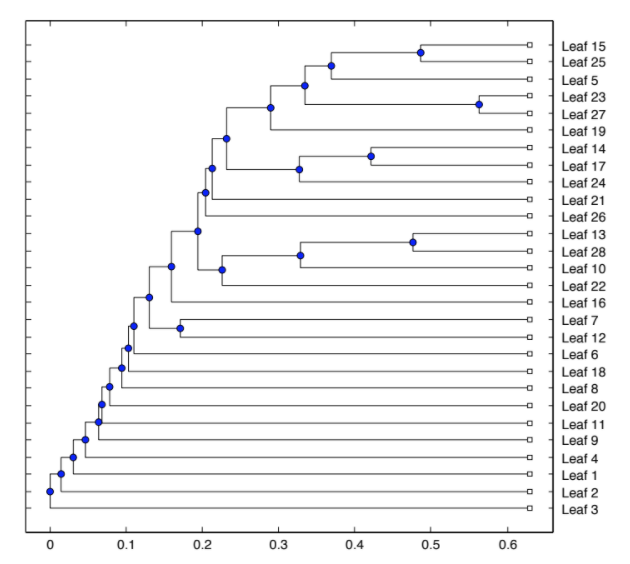}
		\caption{\small{Phylogenetic tree generated for the DroidKungFu family.
Each leaf in the graph denotes a malware sample in DroidKungFu
family, where leaf nodes (1--4) belong to droidkungfu.ab, (5--9)
belong to droidkungfu.aw, (10) belongs to droidkungfu.bb, (11--
12) belong to droidkungfu.bl, (13--15) belong to droidkungfu.c,
(16--22) belong to droidkungfu.g, (23--28) belong to droidkungfu.
m.}}
	\label{fig:phy}
\end{figure}

$\\$
Existing techniques on adversarial malware generation employ feature mutations based on feature vectors extracted from malware. However, most (if not all) of these techniques suffer from a common limitation: feasibility of these attacks is unknown. The synthesized mutations may break the inherent constraints posed by code structures of the malware, causing either crashes or malfunctioning of malicious payloads. To address the limitation, Yang et. al. ~\cite{yang2017ACSAC} present Malware Recomposition Variation (MRV), an approach that conducts semantic analysis of existing malware to systematically construct new malware variants for malware detectors to test and strengthen their detection signatures/models. In particular, we use two variation strategies (i.e., malware evolution attack and malware confusion attack) following structures of existing malware to enhance feasibility of the attacks. Upon the given malware, we conduct semantic-feature mutation analysis and phylogenetic analysis to synthesize mutation strategies. Based on these strategies, we perform program transplantation to automatically mutate malware byte-code to generate new malware variants. We evaluate our MRV approach on actual malware variants, and our empirical evaluation on 1,935 Android benign apps and 1,917 malware shows that MRV produces malware variants that can have high likelihood to evade detection while still retaining their malicious behaviors. We also propose and evaluate three defense mechanisms to counter MRV.

Fig.\ref{fig:malware} gives an overview of the technique we used for malicious app detection. The major differences between our work and the existing works are:

{\bf Mutated samples} We mutate the malware features from the original feature values to the ones that are less differentiable for malware detection~\cite{DBLP:conf/kdd/KongY13}. The newly generated samples are fed into the classification model again for building the discriminant classifier. 

{\bf Evolved samples} We mimic and automate the evolution of malware based on the insight that the evolution process of malware reflects the strategies employed by malware authors to achieve a malicious purpose while evading detection. Fig.\ref{fig:phy} gives an example of phylogenetic tree generated for the DroidKungFu family. The newly generated samples are fed into the classification model again for building the discriminant classifier. 

{\bf Adversarial samples} We explore the malware features to identify much more blind spots of existing detection, and generate the adversarial samples from benign apps that are actually labeled as malicious ones. The newly generated adversarial samples are fed into the classification model again for building the discriminant classifier. 

Our work also has strong connections with adversarial learning~\cite{NIPS2016_6142}.  To make the discriminant classifier more robust, actually we generate the synthetic apps that fools the discriminator into accepting it as the true apps.  Similar to Generative adversarial network (GAN)~\cite{NIPS2014_5423}, these adversarial and mutated/evolved sample generation process is like a generative process, heavily relying on the feature model used in malware detection, instead of the ``random noises'' as in standard GAN.  The generative process actually enforces ``data augmentation'' operations, which is a key strategy used in deep learning process.

Semi-supervised learning is applied for automatic generation of Android security policies in~\cite{DBLP:conf/uss/WangERZNXZA15}. However, it suffers from the inherit limitation of semi-supervised learning~\cite{DBLP:conf/infocom/KongY14}: the algorithm performance degrades significantly at the situation when the mislabeled samples tend to propagate the errors instead of ``belief'' along the similarity measurement path computed from domain knowledge or underlying data manifold information, which is well known in machine learning community.

\subsubsection{App maturity rating}

\begin{figure}[t]
	\centering
	\includegraphics[height=1.8in,width=0.3\textwidth]{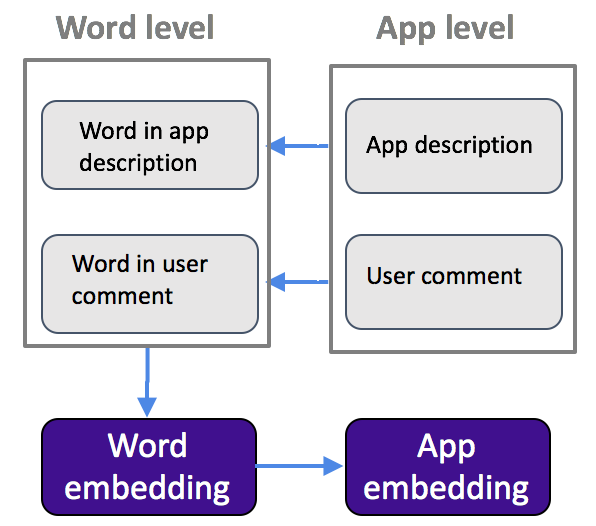}
		\caption{\small{App maturity rating: from word embedding to app embedding.}
	}
	\label{fig:app-rating}
\end{figure}

$\\$ App maturity rating concerns how to protect children from inappropriate content in mobile apps. Apps may contain sexual, violence and drug usage in their content. Therefore, mobile platforms provide rating policies to label the maturity levels of Apps and the reasons why an App has a given maturity level, which enables parents to select maturity-appropriate  Apps for their children. However, existing approaches to implement these maturity rating policies are either costly (because of expensive manual labeling) or inaccurate (because of no centralized controls). In this work~\cite{DBLP:conf/cikm/HuLGKJ15} we aim to design and build a machine learning framework to automatically predict maturity levels for mobile Apps and the associated reasons with a high accuracy and a low cost using machine learning (and deep learning) techniques.

Specifically, we extract novel features from App descriptions by leveraging the semantic embedding of words (a.k.a word embedding) to automatically capture the semantic similarity between words and adapt Support Vector Machine to capture label correlations with Pearson correlation in a multi-label classification setting. In particular, in embedding step, given a sequence of training words $w_1$ , $w_2$ , $w_3$, $\cdots$, $w_T$ , the skip-gram model~\cite{DBLP:conf/nips/MikolovSCCD13} is used to maximize the average log likelihood, {\it i.e.,} %probability  given representation v using:
\begin{eqnarray}
\max \frac{1}{T} \sum_{t=1}^T  \sum_{-c \leq j \leq c, j \neq 0} log \Pr(w_{t+j}|w_j), 
\end{eqnarray}
where c is the size of training context. and $\Pr(w_{t+j}|w_j)$ is the probability of occurrence of $w_{t+j}$ given $w_t$ which is usually defined using a softmax function, {\it i.e.,}
\begin{eqnarray}
\Pr(w_O|w_I) =  \frac {\exp\Big((\vv'_{w_O})^{\top} \vv_{w_{I}}\Big)}  {\sum_{o=1}^W \exp\Big((\vv'_{w_o})^{\top} \vv_{w_{I}}\Big)}
\end{eqnarray}
where $\vv_{w_I}$ and $\vv'_{w_O}$ are the ``input'' and ``output'' word embedding, and $W$ is the number of words in vocabulary.

In app maturity rating application, all app descriptions and textual comments are pre-processed as word-embedding. Then they can be fed into the embedding model perfectly to learn the app embedding.  The key idea is to aggregate the semantics from word level to app level, which actually leverages the semantics from \emph{word embedding} to \emph{App embedding}.  Fig.\ref{fig:app-rating} shows the flowchart of design. Essentially, the framework infers the app maturity using the following logics:
%\begin{eqnarray}

\fbox{%
  \parbox{0.45\textwidth}{%
    \begin{flushleft}
      App description and User comment $\rightarrow$ word embedding  \\
$\rightarrow$ App level embedding $\rightarrow$ Predictive model  
\\
$\rightarrow$  Maturity contents labeling $\rightarrow$  maturity level 
    \end{flushleft}
  }%
}

In experiment, we evaluate our approach and various baseline methods using datasets that we collected from both Apple store and Google Play.  We demonstrate that, with only App descriptions, our approach already achieves 85\% Precision for predicting mature contents and 79\% Precision for predicting maturity levels, which substantially outperforms baseline methods.

\subsection{Marketing for mobile app User Growth}

Marketing the app is one of the most common ways used for driving user growth and improving user engagement. In this section, we will discuss the business intelligence techniques used for attracting new users, promoting the retention of users and improving ad campaign effectiveness.

\subsubsection{Attract more users}

$\\$ There are several typical ways used for attracting new users in business intelligence: 
\begin{itemize}
\item {\bf Concentrating on user experience}:  the best way to build customer relationships is to delight the customers. In a simple word, leave everyone who uses the app feeling good. 
\item {\bf Cross promotion} One need to target people in different segments. Segment the potential customers and figure out how likely one can bring the customer to daily active users (DAU) or monthly active users (MAU).  Targeting model is widely adopted in Facebook, Google and Yahoo business to targeting the look-alike users.  
\item {\bf Develop a Content/functionality Strategy} The app is desired to provide the interesting contents and functionalities that will hit the target audience at the right places. The content and functionality features should cater to targeting's requirement. 
This process is usually optimized using A/B tests to generate even more virality. The ``viral factor'' can be used to measure how effectively the new feature can attract new customers, which is widely adopted in Facebook, Instagram, Snapchat, etc. 
\item {\bf Media exposure} Tell a great story to the media and spend money on different publishers in ad campaigns. The return on investment can be measured using campaign effectiveness analysis. 
\item {\bf Purchasing more traffics} At user-level, different coupons, rewards and discounts are exciting ways to attract new customers. At business side, jointly work with the major internet service providers in bundled sales approach will promote the user growth\footnote{\url{https://hbr.org/2016/02/every-company-needs-a-growth-manager}}.  This is also known as \emph{paid acquisition}.
\item {\bf Search engine optimization} Promote the product using the search engine from creating contents that are more favorable by search engine. The content includes Q\&A, articles, long-form reviews, etc. The goal is to increase the page-views from attracting the new users after searching.  
\end{itemize}

In practice, all these different scenarios can be combined together to promote user growth based on the budget limit. The mathematic optimization is easy to be obtained before reaching the ceiling on saturation, {\it i.e.,} 
\begin{eqnarray}
&&  \max_{\theta, \Theta}  \frac{\Delta \text{\#User} (\Theta)} {\Delta{ \text{Money\_Spent}(\theta)}},  \nonumber \\
&&  s.t. \text{\#User} (\Theta) \leq  \text{Market\_capacity}
\end{eqnarray}
where $\text{\#User} (\Theta)$ and $ \text{Money\_Spent}(\theta)$ denote the number of users and current money spend in the marketplace, and $\Theta, \theta$ are parameters, respectively. In real world, the optimization is performed either empirically from history experience or based on the forecasting results using machine learning and statistical models.  Although the model itself can be highly biased due to the censored and noisy data observations, we believe robust statistical modeling is still an effective and automated way for internet marketing compared to empirical analysis (or simple cohort analysis) from prior knowledge, {\it i.e.,}
\begin{eqnarray}
\Theta = argmin_{\Theta} \mathcal{L} \Big( \text{User number},  \hat{f}( \text{user feature set, app feature set}, \Theta) \Big)  \nonumber \\
\theta = argmin_{\theta} \mathcal{L} \Big( \text{Money spent},  \hat{g}(\text{user feature set, app feature set}, \theta) \Big), \nonumber 
\end{eqnarray}
where $\mathcal{L}(x,y)$ is the loss function (e.g.,  least square loss or cross-entropy loss) that captures the difference between $x$ and $y$, functions $f(.)$ and $g(.)$ can be learned using machine learning/deep learning model in distributed big data science environment.

\subsubsection{Improve user retention}
%%%===
%%%====
\begin{figure}[t]
	\centering
	\includegraphics[height=1.8in,width=0.4\textwidth]{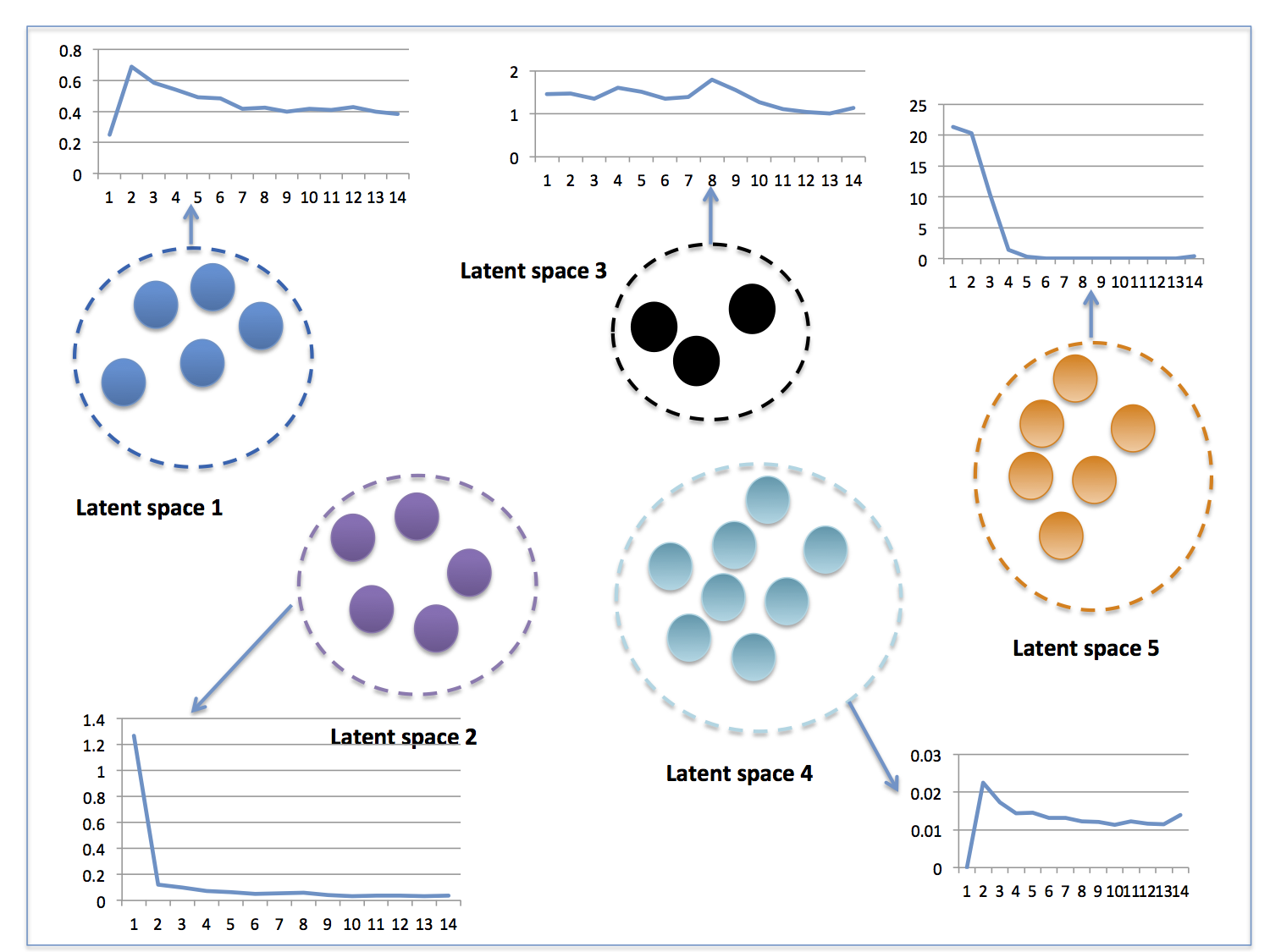}
		\caption{\small{User segmentation (cohort) based on time-varying features.}
	}
	\label{fig:cohort}
\end{figure}
%%%===
%%%====
$\\$
A retention analysis allows one to see the numbers like this:
\begin{itemize}
\item What percentage users are coming back after a week?
\item What percentage of users are paying after a month?
\item How long users stick around for a week or for a month?  
\item Did the new feature released last month increase retention or degrade it?
\item Why some users churn and others do not?
\item How to form growth hypotheses based on quantitative data?
\end{itemize}
Essentially, retention analysis tells a compelling story about ``users doing A in a period" more than ``number of users doing A".  
In cohort analysis, users are grouped into different segmentation (e.g., based on sign-up date)  for behavioral analytics. For example, one can easily observe the percentage of the accounts that still use the service in the following weeks since they signed up at different time. Also, stickiness is a measure of engagement, which looks at how many times a user performs a particular action in a weekly or monthly interval.
In practice, after setting the analytic goals (such as the number of weeks and the cohort of users one wants to track), one can easily write funnel queries or use Google Analytics (or other tools) to achieve the goal. 

Retention analysis can help people to figure out how many people we have lost and drive the insight from such analysis. The numbers shown in retention analysis is not the final goal, but the insight and solutions to improve user retentions are what we really want.  Therefore, it is necessary to figure out the relationship between user actions and retention one you can identify the main behaviors that are correlated with long-term use. The most widely used correlation measurement between retention and an action is {\bf pearson correction}, {\it i.e.,} 
\begin{eqnarray}
r = \frac{\sum_i (x_i - \bar{x}) (y_i -\bar{y})}{ \sqrt{\sum_i (x_i - \bar{x})^2}     \sqrt{\sum_i (y_i - \bar{y})^2}},
\end{eqnarray}
where $x_i$ is sample of retention observations and $y_i$ is sample of user action observations.   For example, for facebook users, 
users who already added 7 friends within the first 10 days after they used facebook are more likely to continue to use Facebook long-term while users who added less than 7 friends are more likely to churn out. For this situation, the user action of ``adding 7 friends in a certain timeframe"  is highly correlated with retention. Please keep in mind that ``correlation'' does not mean ``{\bf causality}''. The user action may not be responsible for the retention. Further causality analysis is needed to connect the cause (i.e., user action) with the effect (i.e., user retention). Fig.\ref{fig:cohort} shows the user segmentation result based on user behaviors.

Further we clarity the relations between retention rate, survival function~\cite{miller-survival}, hazard function (or hazard rate) %from science perspective because these are 
widely used in statistic data science . 

{\bf Hazard Rate} is usually defined as the probability $\lambda_i$ of object that does not survive in the $i$-th time interval ($t_i - t_{i-1}$).

{\bf Retention Rate} is defined as the probability (=$1-\lambda_i$) of object that survives in the $i$-th time interval $(t_i-t_{i-1})$. 

{\bf Survival Function} $S(t_i)$ is defined as the probability of object that survives up to $t_i$. 

Essentially we have: 
\begin{lemma}
Retention rate = 1 - Hazard Rate.  Survival function $S({t_i})$ is given by: 
\end{lemma}
\begin{eqnarray}
\text{Kaplan-Meier Method}:   \quad \quad      S(t_i) = \prod_i (1 - \lambda_i) \\
\text{Nelson-Aalen Method}:   \quad \quad      S(t_i) = \exp{(-\sum_i \lambda_i)}
\end{eqnarray}

\subsubsection{Marketing funnel and  campaign effectiveness}

\begin{figure}[t]
	\centering
	\includegraphics[height=1.6in,width=0.4\textwidth]{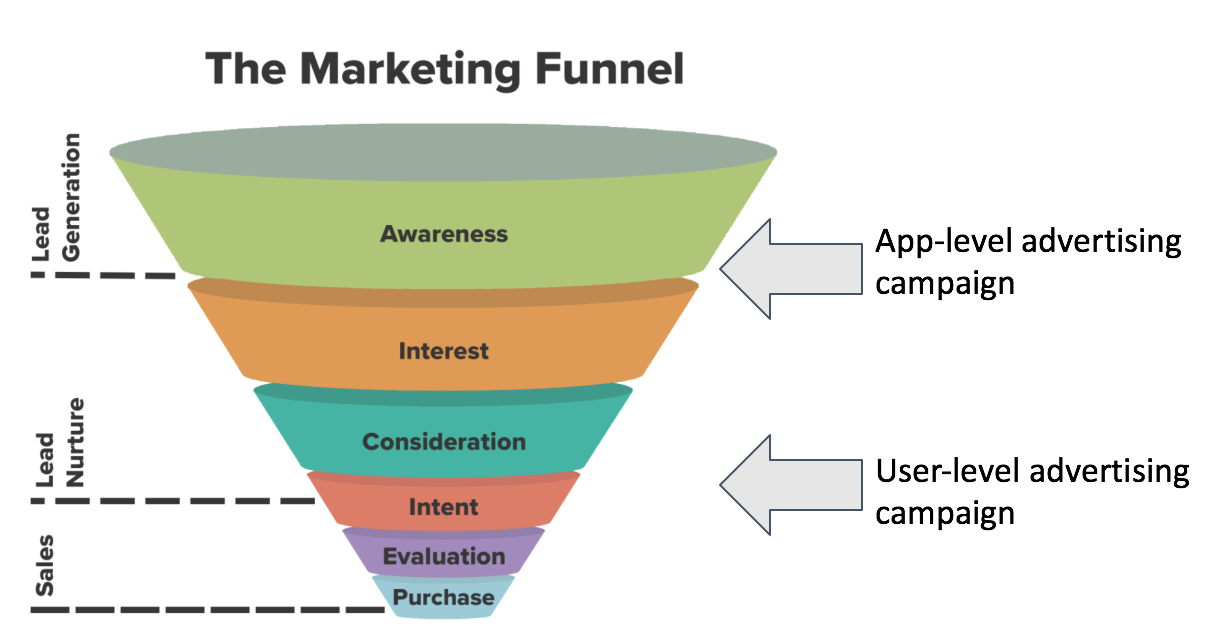}
		\caption{\small{Marketing funnel (from standard marketing science book). The advertising campaigns are started to target different customer cohort. The app-level campaign is generally initiated for brand-level advertising, while user-level advertising is started for lifting up the direct response from users.}
	}
	\label{fig:marketing}
\end{figure}
%%%%==============
%%%%==============
$\\$ 
The marketing funnel is widely adopted in marketing research for understanding the chance of turning leads into customers from marketing and sales perspective.  The general idea is similar to a \emph{funnel}, {\it i.e.}, marketers first give very broad net in order to capture the potential customers as many as possible, and then slowly nurture prospective customers through the purchasing (or conversion) decision, by narrowing down the candidate pools in each stage of the funnel. Fig. shows the different stages in marketing funnel,  {\it i.e.,}

\fbox{%
  \parbox{0.45\textwidth}{%
    \begin{flushleft}
     Awareness $\rightarrow$ interest $\rightarrow$ consideration  \\
$\rightarrow$ intent  $\rightarrow$  evaluation $\rightarrow$  purchase.
    \end{flushleft}
  }%
}

\begin{table}
\begin{tabular}{c|c|c}  
\hline
Campaign Name & Targeting audience & Rewards \\ 
\hline   \hline
Appolo & first time user & New York travel \\  
\hline      
Light burn & MAU &  iphone 8 \\ 
\hline
Fantastic day & Infrequent users & Dogfood \\ 
\hline 
\end{tabular}
\caption{Campaign examples (only for demonstration purpose).}
\label{tbl:campaign}
\end{table}

In practice, at different stages of marketing funnel, the promotion campaign would be different based on targeting customers. For example, at ``awareness'' stage, the campaign goal is to discover the customer net via establishing the trust and thoughts with events, advertising, trade shows, blog posts, infographics, etc. Therefore, the campaign effectiveness measurement should be defined to measure the reaches of customers at different information sources.  At ``evaluation'' stage, one should convince the buyers to make a final decision, then the purchase rate will be an important factor to evaluate the campaign effectiveness.

In fact, marketing campaign can be performed at different stages of funnels for B2C or B2B business. As illustrated before, the campaign effectiveness analysis actually depends on the targeting customers at different stages of funnel, which can be viewed as external variables to \emph{personalize} the campaign effectiveness analysis.  We show the \emph{generalized} version of campaign effectiveness analysis, where personalized version can be obtained correspondingly by incorporating the external variable. 

We show how to compute the campaign effectiveness using generalized analysis. Here we give several examples of targeting campaigns in Table.\ref{tbl:campaign}.

Let $Y_{1i}$   and $Y_{0i}$  be potential benefits (e.g., bonus, rewards, credit, discount, etc) for individual  $i$ when $i$ receives the treatment or does not  receive treatment, respectively.    The fundamental problem of making a \emph{casual inference} is how to  reconstruct the  results that  are not observed. For each individual $i$, what if individual $i$ does not receive treatment?  Basically, we do analysis at aggregation-level using average treatment effect (ATE)  analysis and average treatment  effect on the treated group (ATT) analysis.  ATE is defined as:
\begin{eqnarray}
ATE = E(Y_{1i} |T_{i} = 1, 0) - E(Y_{0i} |T_i  = 1, 0),
\end{eqnarray}
where $E(.)$ represents the expectations on the aggregation level, and $T_i$ denotes the treatment with value $1$ for the treated group while value $0$ for the control group, i.e., the average effect that would be observed in the treated and control group received treatment, compared with if none in both groups received treatment.  Correspondingly,
\begin{eqnarray}
 ATT = E(Y_{1i} |T_i  = 1)  - E(Y_{0i} |T_i  = 1)
\end{eqnarray}
which refers to the average differences if the treated group received treatment if none of these in the treated group received treatment. Then the propensity score~\cite{Mccaffrey04propensityscore} is defined as the conditional probability of receiving a treatment given pre-treatment characteristics:
\begin{eqnarray}
\Pr(X) = \Pr(T=1| X) = E(T|X),
\end{eqnarray}
where $T=\{0,1\}$ is the indicator of exposure to treatment and $X$ is the multi-dimensional representation of pre-treatment characteristics. 
The key idea is that treated and control units should
be on average observationally identical (a.k.a \emph{balancing hypothesis}).  In other words,  for a given propensity score, exposure to treatment is random and therefore any standard probability model can be used to
estimate the propensity score, e.g., 
\begin{eqnarray}
\Pr(T_i =1|X_i) = \phi(h(X_i)),
\end{eqnarray}
where $\phi$ denotes the normal cumulative distribution function, and $h(X_i)$ is a function of covariates with linear and higher order terms. Please keep in mind the choice of estimate of propensity score of function $h(.)$ must satisfy the balancing hypothesis.

\subsection{Mobile App User engagement}
%%%%==============
%%%%==============
\begin{figure}[t]
	\centering
	\includegraphics[height=2.5in,width=0.45\textwidth]{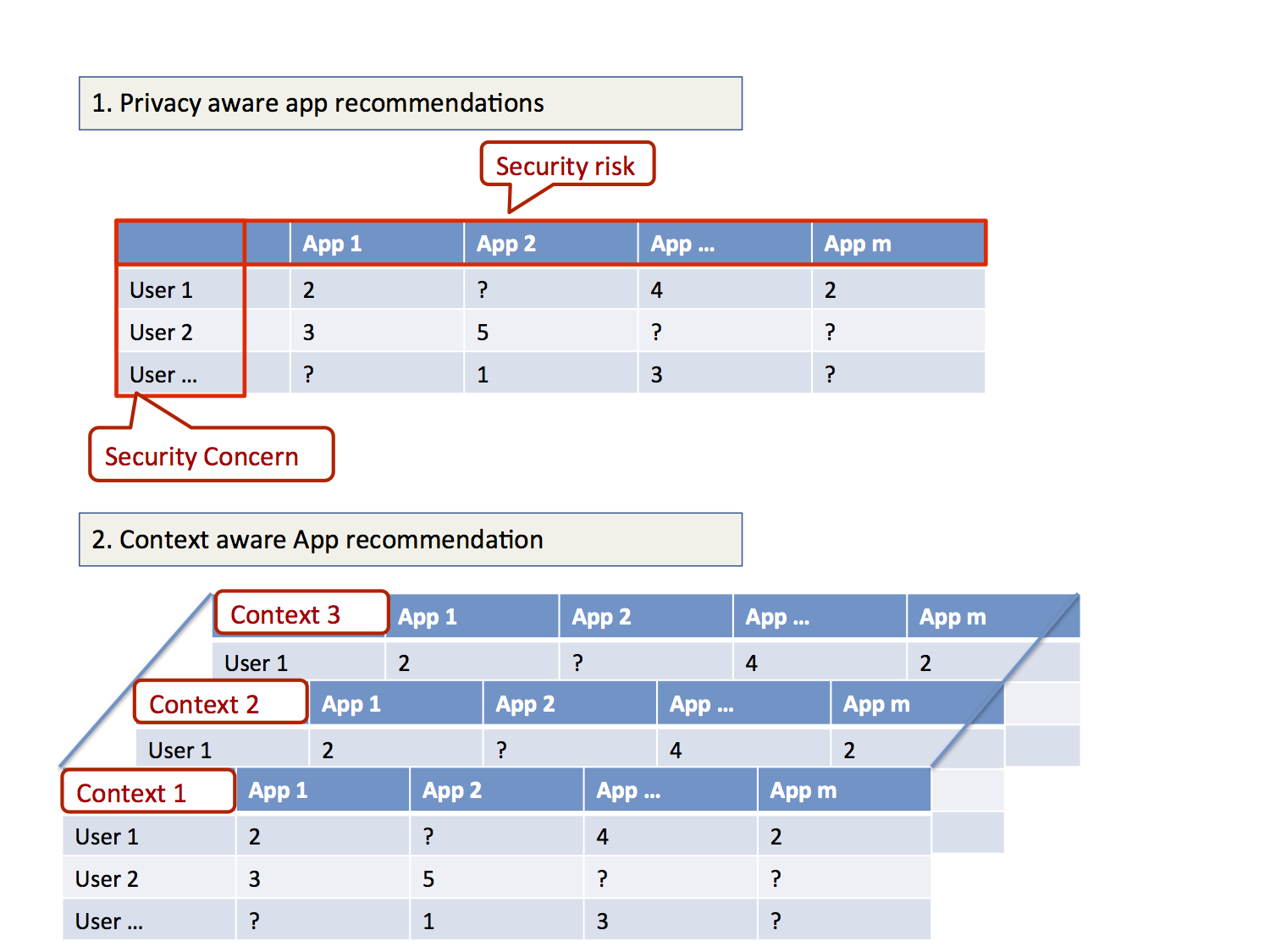}
		\caption{\small{Upper panel: privacy aware app recommendation; Lower panel: context aware app recommendation.}
	}
	\label{fig:context}
\end{figure}
%%%%==============
%%%%==============
In this section, we discuss how to improve the user engagement using user modeling, profiling and recommendation techniques. As the beginning of June 2014, App Store had 1.2 million Apps and a cumulative of 75 billion downloads.  Therefore it is urgent to develop effective personalized App recommendation systems. In particular, we discuss how to do \emph{privacy aware app recommendation} and \emph{context aware app recommendation} for mobile users.  Recommendation is useful and helpful since it can capture users' preference and interest, and it strongly connects with targeting in technology although differs in business logic.  Fig.\ref{fig:context} gives an overview of the app recommendation introduced in this section. 

 \subsubsection{Privacy aware app recommendation}

$\\$
Recent years have witnessed a rapid adoption of mobile devices and a dramatic proliferation of mobile applications (Apps for brevity).  However, the large number of mobile Apps makes it difficult for users to locate relevant Apps. Therefore, recommending Apps becomes an urgent  task.  Traditional recommendation approaches focus on learning the interest of a user and the functionality  of an item (e.g., an App) from a set of user- item ratings, and they recommend an item to a user if the item's functionality  well matches the user interest.  However, Apps could have privileges to access a user's sensitive resources (e.g., contact, message,  and location).  As a result, a user chooses an App not only because of its functionality, but also because it respects the users' privacy preference. To the best of our knowledge, this work presents the first systematic study on incorporating both interest-functionality interactions and user privacy preferences to perform personalized App recommendations~\cite{DBLP:conf/wsdm/LiuKCGJX15}. Specifically, we first construct a new model to capture the trade-off between functionality and user privacy preference. In particular, in this work, it leverages the state-of-the-art Poisson factorization technique and optimizes the objective
\begin{eqnarray}
&& \max_{\uu, \vv} \Pr(y_{ij}| \uu_i, \vv_j, \pp_s), \nonumber \\
&& = Poisson(y_{ij}, \uu_i (\vv_j + \lambda \sum_{s \in \Sigma_j} \pp_s)), 
\end{eqnarray}
where $y_{ij}$  is the rating score for a particular user  $i$ for app $j$,  $\uu_i$  is the user $i$ latent factor,  $\vv_j$  is the app $j$ latent factor and $\pp_s$ is app privacy latent factor with respect to app $j$. Then we crawled a real-world dataset (16, 344 users,
6, 157 Apps,  and 263, 054 ratings) from Google Play and use it to comprehensively evaluate our model and previous methods.  We find that  our method consistently  and substantially outperforms  the state-of-the-art  approaches,  which implies the importance of user privacy preference  on personalized App recommendations. Moreover, we explore the impact of different levels of privacy information on the performances of our method, which gives us insights on what resources are more likely to be treated as private by users and influence user behaviors at selecting Apps.
%%%===============
%%%===============
\begin{table}
\begin{tabular}{c|c|c}  
\hline
User & Location Semantics & Recommended Service \\ 
\hline   \hline
John & Safeway & Apple Pay or Chase Pay \\  
\hline      
Damao &Bank &  Get a free coffee \\ 
\hline
Amy & Mall & Use Banana Coupon \\ 
\hline 
\end{tabular}
\caption{Use cases of context aware recommendation.}
\label{tbl:userec}
\end{table}
%%%===============
%%%===============
\subsubsection{Context aware app recommendation}
$\\$
In many practical applications, in practice, the recommendation depends on context.  Here ``context'' is a very generic concept that can denote location, gender, age, interest or other segmentations. In other words, recommendation is performed on different buckets based on an attribute or a combination of a group of attributes. Similar to app recommendation, we solve this problem using tensor bilinear factorization technique~\cite{kong-context}. In particular, we solve the following problem:
\begin{eqnarray}
&& \max_{\UU, \VV, \PP} \Pr(\XX_{ijk}| \UU_{ir}, \VV_{js}, \PP_{kt}) \nonumber \\
&& = Poisson(\XX_{ijk}, \UU_{ir}\VV_{jr} + \UU_{it}\PP_{kt} + \VV_{js} \PP_{ks})
\end{eqnarray}
where $X_{ijk}$  is the rating score for a particular user $i$ for app $j$ in context $k$, $\UU_{i:}$  is the user $i$ latent  factor, $\VV_{j:}$ is the app $j$ latent factor and $\PP_{k:}$  is app context latent factor for context $k$. Similarly this framework can be easily extended for generating the context aware service recommendations. The use cases are shown in Table.~\ref{tbl:userec}.

{\bf Lessons Learned.}  In this section, we use ``app'' as demonstrated examples for recommendation purpose. Our approach can be easily adapted for recommendations on purchase and others.

\subsection{Mobile App Monetization}
	
The basic idea is how to make more money from mobile apps. Therefore, this section includes \emph{advertisement serving,  advertisement targeting, bid optimization and pricing.}  Although the technology has been graduated into product, this section is intentionally left blank until our research works are published.  %graduated into 

\section{Device-AI solution} 

In this section, we provide device-AI solutions, which aims to protect image privacy on mobile devices.

%%%%==============
%%%%==============
\begin{figure}[t]
	\centering
	\includegraphics[height=1in,width=0.45\textwidth]{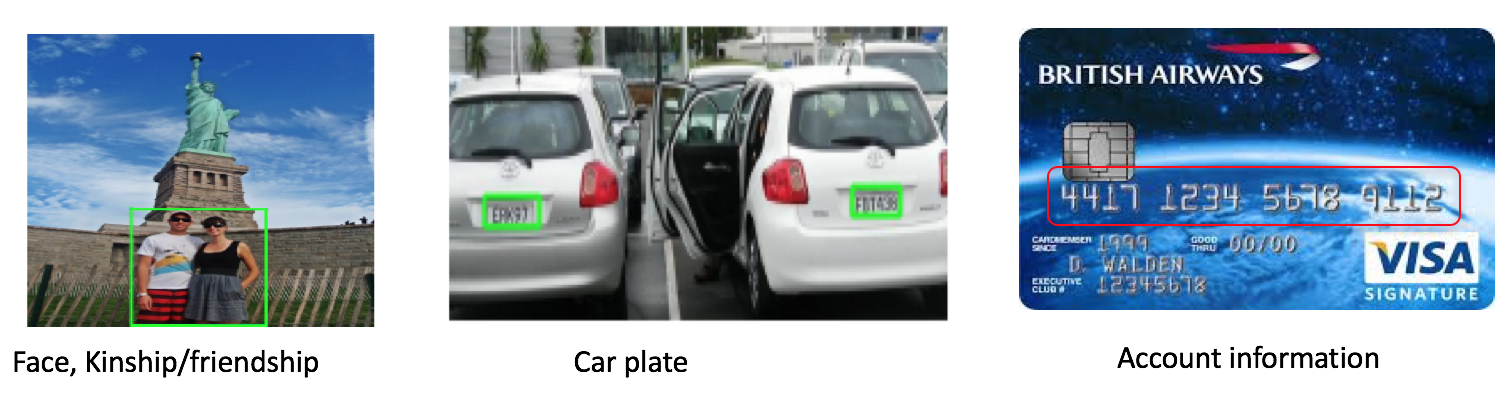}
		\caption{\small{Private photo privacy protection. The sensitive regions are marked using bounding box.}
	}
	\label{fig:photo}
\end{figure}
%%%%==============
%%%%==============
\begin{figure}[t]
	\centering
	\includegraphics[height=0.4in,width=0.45\textwidth]{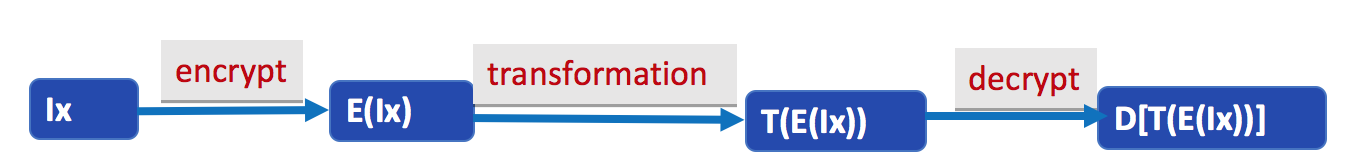}
		\caption{\small{Private photo privacy protection pipeline: original image $I_x$ is encrypted to $E(Ix)$, after transformation it becomes 
		$T(E(Ix))$, and finally after decryption operations it is in the form of $D[T(E(Ix))]$.}
	}
	\label{fig:trans}
\end{figure}

\subsection{Image Recognition and Image Privacy on Mobile Devices}
$\\$	
Every second, nearly 4,000 photos uploaded to Facebook, around 4,600 photos exchanged through Snapchat Photos are uploaded, saved and shared on cloud, e.g., centralized photo sharing platforms (PSPs). In cloud side, sensitive regions are exposed to  public,  etc.  What is  the  security  and  privacy risk?   Photo owners worry about privacy leakage on cloud/PSPs, also Cloud/PSPs may access and process user photos without  explicitly  asking for user agreement and share the unprotected photos. In this work we propose \emph{image perturbation technique} to protect the image privacy.  Ideally, given encryption function $E(.)$, transformation function $T (.)$, the goal is to find decryption function $D(.)$, such that for an image $I_X$ , it exist
\begin{eqnarray}
D\Big[T(E(I_X))\Big] = T(X)
\end{eqnarray}
Our system design is guided by the following theorem. Let $P (.)$  be the image perturbation technique, we have:
\begin{theorem}
Using Image perturbation technique $E= P(.)$  for ``encryption'' of photos, it can exactly ``decrypt" the photos and recover the original one, {\it i.e.,}
\begin{eqnarray}
D\Big[T(E(I_X))\Big] = T(X)
\end{eqnarray}
where $D = f(T, E)$ can be easily calculated given function $E(.)  = P(.)$  and $T(.)$. 
\end{theorem}
Note that ``crypto'' based technique (including symmetric and public key encryption) may not work since it is not compatible with transformation $T (.)$ and also $D$ is impossible to be computed given $E(.)$  and $T (.)$,  and therefore $X$ cannot be recovered (see Fig.\ref{fig:trans}).  For exactly the same reason, differential privacy~\cite{DBLP:series/synthesis/2016Li} added Laplacian noises to the image, which is, in fact, irreversible although privacy preserving. Finally one cannot recover anything given image transformation $T (.)$ as well.

\begin{figure}[t]
	\centering
	\includegraphics[height=1in,width=0.45\textwidth]{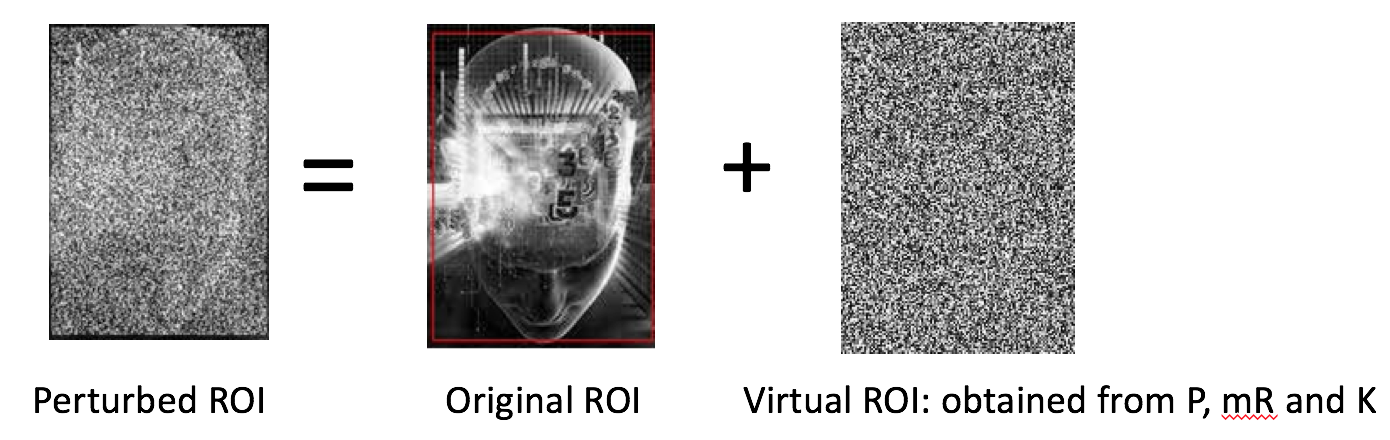}
		\caption{\small{Image perturbation process: the perturbed image is equal to the original image plus the perturbed parameters. The perturbed image is stored on cloud, while the original image is collected from the user. }
	}
	\label{fig:procedure}
\end{figure}

In our approach~\cite{DBLP:conf/dsn/HeLKBWJK16}, it supports different linear transformations (e.g., Rotation, Cropping, Scaling) and also non-linear transformation such as compression. The key idea of our approach is to perturb DC and AC components discriminant in FFT domain, which achieves the same purpose  as crypto but is compatible with different image transformation.  Also, our approach has advantages due to its simple, fast and effective implementation.  In our solution, applying a transformation on perturbed image is equal to applying transformation on original image plus applying transformation on virtual image (generated from perturbation).  An example is shown in Fig.\ref{fig:procedure}.

%%%%==============
%%%%==============
%%%%==============
%%%%==============

\subsection{Deep learning based techniques for protecting image privacy}

%%%%==============
%%%%==============
\begin{figure*}[t]
	\centering
	\includegraphics[height=3in,width=0.9\textwidth]{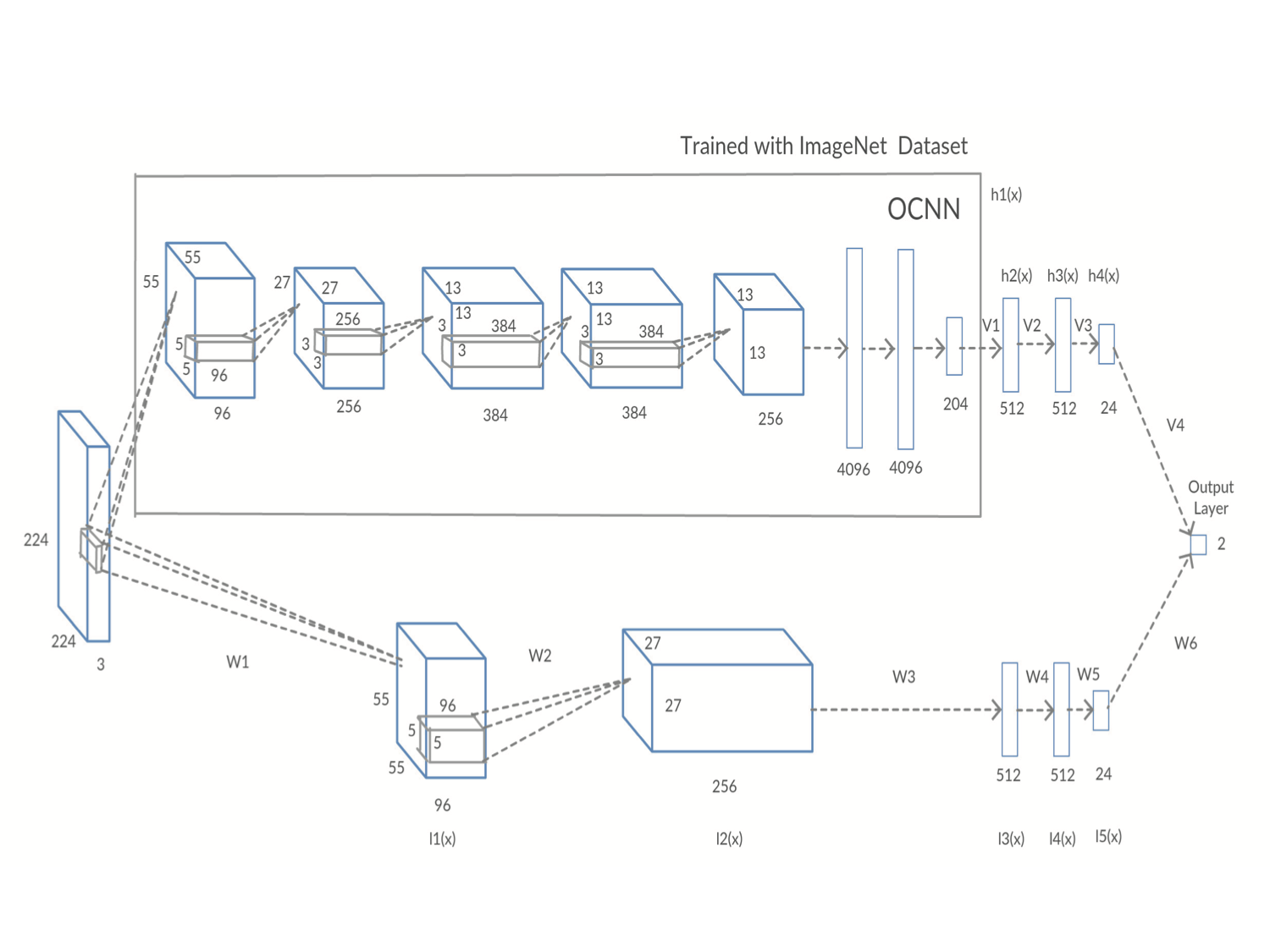}
		\caption{\small{CNN pipeline for privacy detection that consists of two pipelines: (a) object feature learning pipeline (upper panel); (b) convolution feature learning pipeline (lower panel). $h(x)$, $\ell(x)$ are activation functions
		}
	}
	\label{fig:cnn}
\end{figure*}
%%%%==============
%%%%==============
This section provides a deep learning based technique to protect image privacy.  Several examples are shown in Fig.~\ref{fig:photo}.  Photo privacy is a very important problem in the digital age where photos are commonly shared on social networking sites and mobile devices. The main challenge in photo privacy detection is how to generate discriminant features to accurately detect privacy at risk photos.  Existing photo privacy detection works, which rely on low-level vision features, are non-informative to the users regarding what privacy information is leaked from their photos. In this section, we propose a new framework called PrivacyCNH~\cite{DBLP:conf/aaai/TranKJ016} that utilizes hierarchical features which include both object and convolutional features in a deep learning model to detect privacy at risk photos.  In particular, given
the joint deep learning structure (i.e., Alexnet~\cite{NIPS2012_4824})  $\mathcal{V} = \{V^1 , V^2 , V^3 , V^4 \}$
and $\mathcal{W} = \{W^1 , W^2,  \cdots, W^5\}$, the posterior probability
of privacy risk for an image $i$ is given by the sigmoid function using the learned features, {\it i.e.,}
\begin{eqnarray}
&& \Pr(y_i=1| X_i; \mathcal{V}, \mathcal{W}) = \frac{1}{1+\exp{(-z)}}, \\ 
&& z = (V^4_k)^{\top} h_4(X_i) + {(W^6_{\ell})}^{\top} \ell_5(X_i) + \beta, 
\end{eqnarray}
where $V^i$ and $W^j$ are the CNN network structure parameters with $i$ and $j$ indicating the layer number, $k$ indexes the hidden unit in layer $i$, $\ell$ indexes the hidden unit in layer $j$,  $h_i$  and  $\ell_j$  are the activation functions for object  CNN and low-level CNN respectively and $\beta$ is the biased scalar term.

The generation of object features enables our model to better inform the users about the reason why a photo has privacy risk. The combination of convolutional and object features provides a richer model to understand photo privacy from different aspects, thus improving photo privacy detection accuracy. Experimental results demonstrate that the proposed model outperforms the state-of-the-art work and the standard convolutional neural network (CNN)  with convolutional features on photo privacy detection tasks. Fig.~\ref{fig:cnn} demonstrates the pipeline of our method.

\subsection{Model compression on mobile devices}

%%%%==============
%%%%==============
\begin{figure}[t]
	\centering
	\includegraphics[height=2in,width=0.4\textwidth]{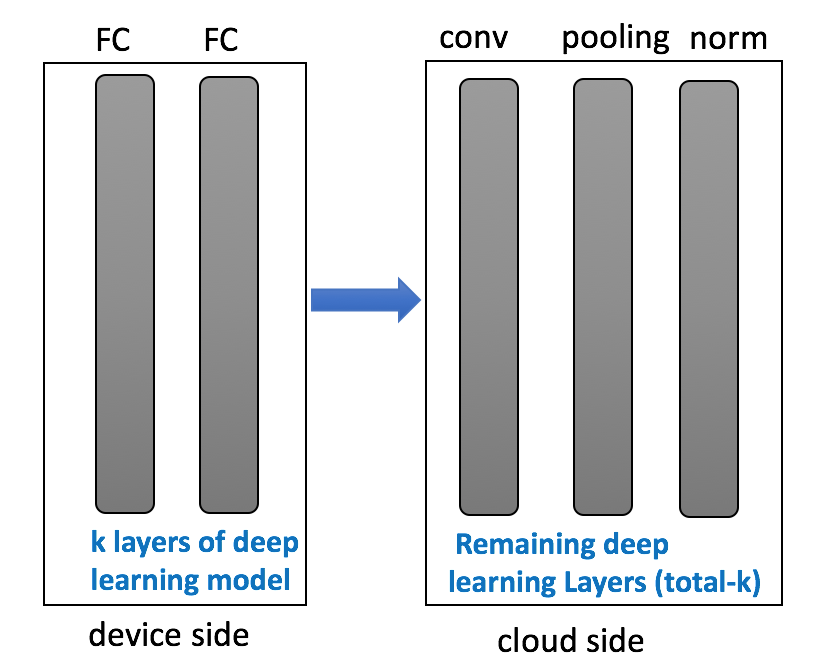}
		\caption{\small{Distributed Deep Learning execution framework.
		}
	}
	\label{fig:ddl}
\end{figure}
%%%%==============
%%%%==============

CNN model has been widely and successfully used in many computer vision tasks, such as object detection, fine-grained image classification, age estimation~\cite{ DBLP:conf/fgr/WangZKCLZ17}, etc. The popularity of mobile phone brings the great convenience to people life due to the existence of many practical and excellent apps. However, to run CNN models (even in testing phrase) for a typical vision task is a luxury for most devices due to the high computational cost and limited memory space and power resources. To accelerate CNN models is highly desirable to facilitate mobile vision applications that highly depends on the performance of CNN models.

Our investigation on AlexNet indicates that  not only full-connected layers and convolution layers consume a lot of time, but also some non-tensor layers (such as Pooling layer and LRN layers) that  do not contain any high-order tensor-type  weight  parameter are also time-consuming. However, current researches focus on rank approximation or parameter compression in fully connected and convolution layers.  Although helpful, the acceleration and compression of non-tensor layers are totally ignored.

To address this limitation, this paper~\cite{li2017deeprebirth} proposes a unified framework to compress CNN models by dismembering non-tensor layers, to simultaneously accelerate the CNN model testing performance with neglect performance degradation. With re-trained new network parameters in ``re-birth'' layers, the functionality of non-tensor layers is equivalently implemented in the new merged layers with significant efficiently improvement. The standard least square error is used to minimize the error function in re-training process where the new parameters are essentially the ``quantized'' old parameters (in some sense). The framework includes both \emph{streaming merge} and \emph{branch  merge} that are able to conduct fast computations easily adapted for current mainstream CNN models and potential new CNN pipelines. In the meantime, in order to run deep learning on mobile devices, we provide an \emph{elastic} approach to run deep learning in a distributed fashion (shown in Fig.~\ref{fig:ddl}). 

{\bf Theoretical Analysis}

The convolution layer transforms the input feature map $X \in \Re^{M \times N \times K} \rightarrow Y \in \Re^{M' \times N' \times K'}$, {\it i.e.} 
\begin{eqnarray}
\label{EQ:conv}
f_{\text{conv}}:    X \mapsto Y, \;\;\;  Y_{i'j'k'} = \sum_{i=1}^{d_k}  \sum_{j=1}^{d_k} \sum_{k=1}^K W_{ijkk'} X_{i+i',j+j',k}   \;\; (1 \leq k' \leq K') \nonumber
\end{eqnarray}
where $K, K'$ are the number of feature map channels, and $M, N; M', N'$ are the size of the images, which is actually 
regular linear convolution by a filter bank, $d_k \times d_k$ is the kernel size and feature map $Y$ is essentially the sum of inner product by traversing along 
different locations with $d_k \times d_k$ kernel (e.g., $d_k = 3$) and the output response  $Y$is obtained by enforcing linear transformation $\WW$ on feature map $X$.

The local response normalization (LRN) layer performs ``lateral inhibition'' based on the fact that the activated neurons will have impact on those neurons in its local input regions. Therefore, it usually performs normalizing over local input regions from $ \Re^{M \times N \times K} \rightarrow  \Re^{M \times N \times K'} $,   {\it i.e.,}
\begin{eqnarray}
\label{EQ:LRN}
f_{\text{LRN}}:   X \mapsto Y,     \;\; \;   Y_{ijk'} = \frac{X_{ijk}}{\left(\kappa + \alpha \sum_{k\in G(k')}  X_{ijk}^2\right)^{\beta}},
\end{eqnarray}
where $G(k) = \left[k - \lfloor \frac{\rho}{2} \rfloor, k + \lceil \frac{\rho}{2} \rceil\right] \cap \{1, 2, \dots, K\}$
is a group of $\rho$ consecutive feature channels in the input map. Clearly, if $\kappa=0, \alpha = 1, \beta =1$, this gives $\ell_2$ normalization.  A batch normalization operation~\cite{icml2015_ioffe15} is usually applied to change the distributions of activations to avoid ``Internal covariate shift''.  During SGD training, each activation of the mini-batch is centered to zero-mean and unit variance where the mean and variance are measured over the whole mini-batch. Then a learned offset $\beta$ and multiplicative factor $\gamma$ are then applied, i.e., given values of $X$ over a mini-batch: 
$\{X^1, X^2, \cdots,X^m\}$, batch normalization projects them into  $\{Y^1, Y^2, \cdots, Y^m\}$ using the following steps: 
\begin{eqnarray}
%f_{\text{BN}}:  \RR^{M \times N \times K} \rightarrow  \RR^{M \times N \times K};   \;\;  X \mapsto Y,\\
\label{EQ:BN}
&& \mu_{\BB} \leftarrow  \frac{1}{m} \sum_{i=1}^m X^i;  \;\;\;  
\sigma^2_{\BB} \leftarrow  \frac{1}{m}  (X^i - \mu_{\BB})^2;  \;\;\;   \nonumber \\
&& \hat{X^i} \leftarrow \frac{X^i - \mu_{\BB}}{ \sqrt{ \sigma^2_{\BB} + \epsilon}};   \;\;\; 
 Y^i \leftarrow \gamma \hat{X^i} + \beta;
\end{eqnarray}
where $\mu_{\BB}$ and $\sigma^2_{\BB}$ are the mean and variance of the data in the mini-batch. 

Then a pooling operator operates on individual feature channels, coalescing nearby feature values into one by the application of a suitable operator such as max pooling from $\Re^{M \times N \times K} \rightarrow  \Re^{M' \times N' \times K'}$, {\it i.e., }  
\begin{eqnarray}
\label{EQ:pooling}
f_{\text{Pooling}}:  X \mapsto Y,    \;\; \;  Y_{ijk} = \max \{ Y_{i'j'k} : i \leq i' < i+p, j \leq j' < j + p \},  \nonumber 
\end{eqnarray}
where $p$ denotes the nearby $p$ regions\footnote{Sum pooling can be similarly done.}. 

To achieve the desired functionality with acceleration, the idea is to find a mapping function $\mathcal{F}:  X \in \Re^{ M \times N \times K} \rightarrow Y \in \Re^{ M'' \times N'' \times K'} $ such that it can get the same feature map value $Y^i$ given the same input feature map $X^i$ for any image $i$.  Recall that convolution operation can be viewed as enforcing linear transformation $W$ on the input feature maps in the fully connected layers, and therefore we aim to build a single convolution operation ($*$) that replaces several non-tensor layers by setting a new optimization goal,  {\it i.e., }   
%\begin{eqnarray}
%X_i \rightarrow Y_i ;  \;\;\;  X_i \in \RR^{ M \times N \times K},   Y^{\text{COM}}_i \in \RR^{ M'' \times N'' \times K'},
%\end{eqnarray}
%Our goal is to build a neural network with only convolution operations, {\it i.e.,}
\begin{eqnarray}
\label{EQ:y_i = Y_com}
\forall i:   \; \; \;   Y^i = Y_{\text{COM}}^i ;   \;\;\;  Y_{\text{COM}}^i  \simeq \hat{W} * X^i + \hat{b};  
\end{eqnarray}
While the type and sequence of functions is usually handcrafted, the parameters $W$ and bias $b$ can be learned from our experiments for solving a least square problem using SGD, {\it i.e.,}
%Clearly,  the optimal solution $\hat{W^*}, \hat{b^*}$ is given by minimizing the following least square loss function, {\it i.e.,}
\begin{eqnarray}
\label{EQ:W*b*}
(\hat{W^*}, \hat{b^*} ) = argmin_{\hat{W},\hat{b}} \sum_i \|  Y_{\text{COM}}^i  -  (\hat{W} * X^i + \hat{b}) \|^2, 
\end{eqnarray}
%Thi can be easily solved using stochastic gradient descend method.   
Note here $\hat{W^*} \in \Re^{d_s \times d_s \times K \times K'}$, the size of which is quite similar to the original convolution steps except new kernel size changes to $d_s \times d_s$ by considering pooling operations.

\section{Cloud and Device-AI interactions}

%%%%==============
%%%%==============
\begin{figure}[t]
	\centering
	\includegraphics[height=2in,width=0.35\textwidth]{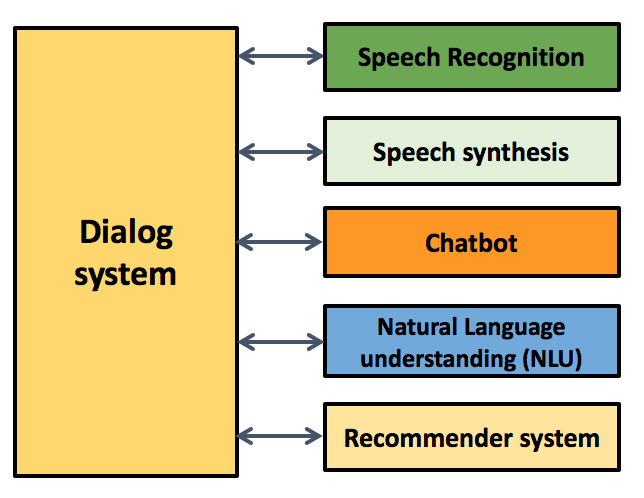}
		\caption{\small{Key component in ``personal intelligent assistant''.}
	}
	\label{fig:alexa}
\end{figure}
%%%%==============
%%%%==============

A typical example is ``Intelligent personal assistant".  The ideal personal assistant can talk to you in speech conversation, understand your words, get information for you and even do something for you (such as write a letter).   The major several products in markets include Amazon Echo\footnote{\small https://en.wikipedia.org/wiki/Amazon\_Echo}, Google Home\footnote{\small https://en.wikipedia.org/wiki/Google\_Home},  Apple's Siri\footnote{\small https://en.wikipedia.org/wiki/Siri},  and Microsoft Cortana\footnote{\small https://en.wikipedia.org/wiki/Cortana}. For example, Amazon Alexa  has been sold more than 25 millions since 2015\footnote{\small https://en.wikipedia.org/wiki/Amazon\_Alexa}. The personal assistant can work in the following 
scenarios:
\begin{itemize}
\item Online chat (such as some instant message app
\item Speech and voice recognition
\item  Taking and uploading images.
\end{itemize}
The personal assistant can provide a wide variety of services, such as providing weather information, playing music from Spotify\footnote{\small https://www.spotify.com/us/}, playing videos, buying items from Amazon, completing the customer service tasks, {\it etc}.  Technically, personal assistant needs six major components to support all its functionalities, i.e., 
\begin{itemize}
\item Speech Recognition
\item Speech Synthesis
\item Natural Language Understanding (NLU)
\item Chatbot
\item Dialog system
\item Recommender System
\end{itemize}
which will be illustrated in detail in the following subsections. 

\subsection{Speech recognition}
Essentially, speech recognition component learns a function $\hat{g}$ which can automatically label the speech signal to its corresponding labels (such as transcript) in a structured-input structured-output way, {\it i.e.,}
\begin{eqnarray}
\hat{g}:   \text{speech signal} \rightarrow  \text{label of signal}
\end{eqnarray}
The goal is to learn function $\hat{g}$ with high accuracy.  $\hat{g}$ can be modeled using HMM, LSTM and other deep learning models. 

The tradition phonetic-based (e.g., HMM-based method)~\cite{Gales:2007:AHM:1373536.1373537} required feature engineer (e.g. n-gram) and separate training component for pronunciation, acoustic and language model. The benefit of the current end-to-end deep learning pipeline essentially jointly learns different components of speech recognizer, which facilitates the deep learning training and deployment on mobile device. 
Attention-based automatic speech recognition models (a.k.a "Listen, Attend and Spell")~\cite{Chorowski:2015:AMS:2969239.2969304} based on deep neural network can literally listen to the acoustic signal, pay attention to different parts of the signal and spell out the transcript one character at a time. 

From product perspective, the automatic speech recognition (ASR) module needs the support of the far-field technology, speaker adaption, noise filtering techniques~\cite{Li:2014:ONA:2687012.2687013} to make the ASR system work in practice.  For example, far-field E (electric) and B (magnetic) field strength decreases inversely with distance from the source, resulting in an inverse-square law for the radiated power intensity of electromagnetic radiation.

%%%%==============
%%%%==============
\begin{figure}[t]
	\centering
	\includegraphics[height=1.5in,width=0.35\textwidth]{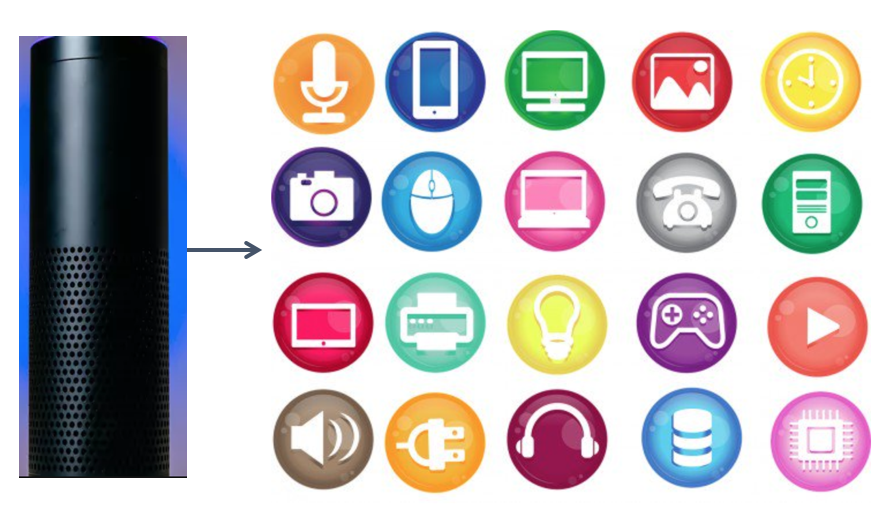}
		\caption{\small{The trigged functionalities provided by the most popular personal assistant: Amazon Alexa. The Alexa and app pictures are obtained from internet. }
	}
	\label{fig:alexfun}
\end{figure}
%%%%==============
%%%%==============

\subsection{Speech Synthesis}

Speech synthesis~\cite{Taylor:2009:TS:1592988} aims to synthesize the human speech from the textual descriptions or from  symbolic linguistic representations. 
A typical text-to-speech (TTS) system (or "engine")~\cite{Dutoit:1997:ITS:262411} has several components, and the learned function $\hat{h}$ can automatically generate the speech accurately given the input text, {\it i.e.,}
\begin{eqnarray}
\hat{h}:   \text{text characters} \rightarrow  \text{speech}
\end{eqnarray}

The front end first performs tokenization and normalization, which converts raw text (including numbers and abbreviations) into the written-out words. Then each word was assigned the phonetic transcription, which divides the text into prosodic units such as clauses, phrases, and sentences.  This actually finishes the text-to-phoneme conversion. After this process, the output is the symbolic linguistic representation consisting of phonetic transcriptions and prosody information.  

The next thing concerns how to convert the symbolic linguistic representation into sound.  The general way is to compute the target prosody (pitch contour, phoneme durations) on the output speech.

It seems the process is very complicated. Fortunately many APIs from big giants (e.g., Apple Siri, AT\&T) have been offered to accelerate the development process.

%xt normalization, pre-processing, or tokenization. The front-end then assigns phonetic transcriptions to each word, and divides and marks the text into prosodic units, like phrases, clauses, and sentences. The process of assigning phonetic transcriptions to words is called text-to-phoneme or grapheme-to-phoneme conversion. Phonetic transcriptions and prosody information together make up the symbolic linguistic representation that is output by the front-end. 

 \subsection{Natural Language understanding (NLU)}
 
 Natural language understanding module processes the natural language input using disassembling and parsing techniques~\cite{Huang:2001:SLP:560905}. Given the utterance, the system needs to identify the proper name, part-of-speech (POS), named entity and finally parses it into the object and predicator sets. Natural language understanding is performed from syntax level, semantic level to pragmatic level of understanding in linguistic analysis, which lays foundation for understanding the sentiment and emotions~\cite{Liang:2006:EDA:1220175.1220271}, uncovering insights from structured and unstructured data.  The three most important features desired for NLU are:
 \begin{itemize}
 \item Proper name identification: identify the ``proper name" from utterance
 \item Part of speech tagging: labelling the part of speech as a category of words that have similar grammatical properties.  For example, in english, the labels are  noun, verb, adjective, adverb, pronoun, preposition, conjunction, interjection, etc. 
 \item Syntactic/semantic parser:  taking input data and building the data structure - in the format of parse tree, such as abstract syntax tree or other hierarchical structure  to represent the input and check the correctness of syntax or semantics. % correct syntax in the process. 
 \end{itemize}
 
 \subsection{Chatbot}
 
Chatbot is a computer program that can conduct the conversation using textual or audio methods with human being. The chatbot simulates how a human would behave in a conversation, which is widely used in a dialogue system or customer service for practical purpose. The simple chatbot uses keyword-based matching for the input, and then queries the answers from database using keyword matching from the prior knowledge database.  Nowadays, more sophisticated natural language processing techniques (e.g., recurrent neural networks/LSTM) are used in current Chatbot systems~\cite{DBLP:journals/corr/abs-1709-02349}.

\subsection{Dialog System}

The dialog system~\cite{DBLP:journals/corr/LowePSP15} intends to have conversations with human like a conversion agent from employing text, speech, graphics, haptics, gestures and others for communication.  

One typical dialog system using reinforcement learning first triggers the language understanding component to understand the use input, and then dialog manager queries the dialog policy, {\it e.g.,} 
\begin{eqnarray}
a = \pi(s)
\end{eqnarray}
where $a$ is the action given the current sate $s$ distribution $\pi(s)$, then it collects the rewards using
\begin{eqnarray}
\mathcal{L} = (s, a, r, s'),
\end{eqnarray}
where $r$ is the reward and $s'$ is the state after transition. The Q function can be induced by applying Q-learning updates over mini batches
such that $Q$-function is optimized using
\begin{eqnarray}
\max Q(s,a),
\end{eqnarray}
before making a decision on the dialog policy. After the dialogue policy is determined, the dialog system manages the general flow of the conversation based on the history and state of the dialog, and produces output using the output generator, including natural language 
 generator and trigger text-to-speech engine (TTS) as well.  When the $Q$-function is learned, to achieve the better performance, using multilayer deep learning (e.g., GoogLeNet~\cite{43022}) networks, we in fact, adopt deep reinforcement learning~\cite{DBLP:journals/corr/LiMRGGJ16} to build the more robust and accurate system.

 \subsection{Recommender System}
 
A typical way is to incorporate more information into both Item features and user feature modeling process, such as the content information of the items (including title, artists, genre, year, etc), user demographics (including income, age, gender), geo-location, social network profile and other relevant features. Given millions of features and data samples, a natural way is using distributed big machine learning framework to derive the corresponding solutions using ``learning to ranking" functions, {\it i.e.,}
\begin{eqnarray}
\hat{f}:  (\text{visit, item title, user income,...}) \rightarrow \text{likelihood of users' like or dislike} \nonumber
\end{eqnarray}
Google implemented this type of recommendation system using deep and wide recommendation~\cite{DBLP:journals/corr/ChengKHSCAACCIA16}. 
 
{\bf Summary} Imagine that one can easily build his own personal assistant, given the far-field recognition technique on device, Amazon cloud service (AWS), state-of-the-art speech recognition and natural language understanding techniques, chatbot and dialog system, recommender techniques. It is the great opportunity for everyone. %sudo find / -name *.awk

\section{Conclusion}
This paper presents our research efforts on mobile data science, which provides a scientific approach to drive innovations on different mobile AI applications on both cloud side and device side. The paper presents very detailed case studies regarding how to apply machine learning and optimization techniques to solve the real-world challenging problem on mobile devices.  We are delivering more intelligent mobile innovations for mobile AI applications.  Stay tuned for our future works.

{\bf Acknowledgement }
The majority of paper contents are based on the author's \emph{published} papers. Thanks for all co-authors who have contributed to this work as listed in references, and I really appreciate their strong support and help.  Any  opinions,  findings  or recommendations expressed in this material  are those of the authors and do not necessarily reflect the views of any company.  Any plagiarism of this work is forbidden (especially for bad guys).

\bibliographystyle{ACM-Reference-Format}
\bibliography{mobileai}

\end{document}